# Mitigation of Delamination of Epitaxial Large-Area Boron Nitride for Semiconductor Processing

Jakub Rogoża, Jakub Iwański, Katarzyna Ludwiczak, Bartosz Furtak, Aleksandra Krystyna Dąbrowska, Mateusz Tokarczyk, Johannes Binder, Andrzej Wysmołek

*Faculty of Physics, University of Warsaw, Pasteura 5, 02-093 Warsaw, Poland*

**Abstract**

Hexagonal boron nitride (hBN) is a promising material for next-generation semiconductor and optoelectronic devices due to its wide bandgap and remarkable optical properties. To apply this material in the semiconductor industry, it is necessary to grow large-area layers on the wafer-scale. For this purpose, chemical vapor deposition methods are highly preferable. However, in the case of epitaxial BN, its fragility and susceptibility to delamination and fold formation during wet processing, such as lithography, present significant challenges to its integration into device fabrication. In this work, we introduce a controlled delamination and redeposition method that effectively prevents the layer from degradation, allowing for multi-step lithographic processes. This approach is applicable to BN layers across a broad thickness range, from tens to hundreds of nanometers, and ensures compatibility with standard photolithographic techniques without compromising the material's intrinsic properties. By addressing key processing challenges, this method paves the way for integrating epitaxial BN into advanced semiconductor and optoelectronic technologies.

**Main**

Among the advanced two-dimensional materials currently under intense investigation, hexagonal boron nitride (hBN) stands out due to its exceptional physical and chemical properties. Its wide bandgap of 5.95 eV[1] makes it an ideal dielectric for use in 2D material-based systems[2–4], enabling the development of next-generation electronic and optoelectronic devices. Beyond its role as a dielectric, hBN's wide bandgap and efficient deep-ultraviolet (DUV) light emission[1,5–7] position it as a key material for emerging DUV optoelectronic applications, such as compact and robust light emitters[8] and detectors for DUV range[9–12].

For industrial applications, the development of high-area, high-quality hBN layers is essential. Chemical vapor deposition (CVD) has emerged as one of the most viable method for growing BN films of varying thickness[13–16]. The importance of using various thicknesses of hBN lies in their distinct applications—thick layers are essential e.g. in optical detectors due to their higher absorption capabilities[17] and for neutron detection[17,18], while thin layers are more useful as functional layers e.g. in stacking structures where different 2D materials interact to achieve novel properties. However, the fragility of epitaxial BN layers and their propensity to delaminate, a process in which the large-area layer is entirely peeled off by the action of liquid, during processes such as lithography[19,20], remain key challenges for scalable manufacturing.

Interestingly, this delamination characteristic, often viewed as a drawback, has proven advantageous in research settings, where liquid-based transfer techniques allow the stacking of hBN layers to achieve more complex structures[21]. These structures can enable BN to act as an anti-oxidation layer or an electric charge-blocking layer [22–25].

Nevertheless, in industrial settings, uncontrolled delamination can hinder the integration of hBN in device fabrication, especially during multi-step processes like lithography. This problem is highlighted in the literature e.g. during development of solid-state based neutron detectors of considerable thickness[17]. Unwanted delamination has also been addressed by other groups, and one proposed approach is to control the diffusion of aluminum atoms into hBN from aluminum nitride (AlN) buffer layers[26]. By adjusting the growth temperature of the AlN buffers, aluminum diffusion can be modulated, thereby achieving either desired lift-off of the h-BN layer or creating robust, mechanically inseparable h-BN layers.

In response to this challenge, we have implemented an original approach based on a controlled delamination technique that intentionally preprocesses hBN layers prior to critical fabrication steps, allowing for precise handling and integration. Our approach is versatile, accommodating hBN layers across a broad range of thicknesses—from thin epitaxial layers (tens of nanometers) to thicker films (hundreds of nanometers). This innovation resolves a technological bottleneck, enabling epitaxial BN to be processed using standard semiconductor technologies.

**Methods**

The samples investigated in this paper were grown in an AIXTRON CCS 3x2" metalorganic vapor phase epitaxy (MOVPE) reactor, using 2''sapphire as a substrate. Ammonia and triethylboron (TEB) served as precursors for nitrogen and boron, respectively. Hydrogen carrier gas was used to introduce the reagents into the reactor. The work presented in this article involved the growth and analysis of multiple samples; however, the discussion focuses on two representative examples that best illustrate the growth of thick and thin epitaxial layers on sapphire. A thick layer (~800 nm) was grown using continuous flow growth (CFG) mode[27], where ammonia and TEB were introduced simultaneously at a temperature of 1200°C, under a pressure of 600 mbar in the reactor, keeping a V/III flow ratio of about 200. The second, thin layer (~30 nm), was grown using a two-stage method described in detail in refs[28,29]. The first stage is to grow a thin buffer layer, which was deposited over 10 minutes in CFG mode. Subsequently, the process proceeds to flow-rate modulation epitaxy (FME) growth conditions, where precursors are alternately supplied to the reactor[28,29], at 1300°C, and under a pressure of 400 mbar. The sample was then cooled to room temperature.

Comprehensive characterization of the samples was performed using different techniques. Scanning electron microscopy (SEM) was carried out using an FEI Helios NanoLab 600 system to examine surface morphology. Fourier-transform infrared (FTIR) spectroscopy was performed with a Thermo Fisher Scientific iS50 system, equipped with a Nicolet Continuum microscope and a 32× infinity-corrected reflective objective (NA 0.65). Each sample was measured in 7 areas of 70×70 $\mu m^2$ with an interdistance of ~5 mm demonstrating great homogeneity across the sample. Raman measurements were performed with Renishaw inVia spectrometers. The material was excited by a λ = 532 nm laser with a power of 30 mW and a 100× objective (NA 0.85).

The optical setup employed for photoconductivity measurements of the epitaxial BN used an EQ-99X LDLS plasma lamp as the light source, capable of operating over the 170–2500 nm range. The divergent beam emitted by the lamp was collimated using a parabolic mirror and subsequently focused by a spherical mirror onto the entrance slit of a Triax 550 monochromator (Horiba), equipped with a 3600 lines grating. The monochromatic light from the monochromator was then directed to the sample. The photoinduced changes of sample resistance were detected using a EG&G 5210 Dual Phase Lock-In Amplifier.

Photolithographic structures were fabricated using a ma-P1215 photoresist (micro resist technology GmbH). The photoresist was deposited via spin coating at 3000 rpm for 30 seconds, resulting in a uniform layer approximately 1.5 μm thick. Exposition was performed with a POLOS μPrinter maskless lithography system. Alignment markers in the shape of "+" were incorporated at the mask corners when necessary to facilitate precise pattern alignment. Following exposure, the photoresist was developed in MF331 solution for 60 seconds, rinsed with deionized water for 10 seconds to remove residual developer, and dried with a stream of compressed nitrogen. To improve mask durability during etching process, the resin was cured in an oven at 100°C for one hour.

Sample etching was conducted using an RF-ICP system (Oxford Instruments Plasmalab 100) with $SF_6$ gas supplied at a flow rate of 40 sccm. The etching process lasted 60 seconds, with RF power set to 20 W and ICP power to 50 W. Post-etching, the mask was removed by immersing the sample in acetone until the resist was fully dissolved. Residual acetone was eliminated through rinsing with deionized water, followed by drying with compressed nitrogen. The sample was subsequently placed in a vacuum oven at 100 °C for one hour to ensure complete drying.

Metal deposition was performed using a Gatan PECS 682 Precision Etching and Coating System. An AuPd layer of 50 nm thickness was sputtered at 10 kV. The lift-off process was carried out by submerging the sample in acetone bath on a planetary mixer. If needed, samples were submerged in ultrasonic bath to remove metal residues.

**Results**

The delamination of epitaxial BN is a well-documented phenomenon[19–21,26,30] with broad applicability, as outlined in the introduction. In our work, we have extensively explored and successfully applied this process in various contexts, including the redeposition of our epitaxial BN onto $SiO_2$ to serve as a substrate for the growth of $MoSe_2$[20] or to transfer BN onto germanium, which can be etched using photocorrosion process, enabling the fabrication of free-standing membranes that have potential applications in enhancing optical signals[21].

The delamination of epitaxial BN has been further investigated, revealing that the thickness of the BN layer strongly influences its adhesion to the substrate. Thicker BN layers tend to delaminate more easily. For layers thicker than 100 nm, complete delamination is likely when exposed to water[19]. In contrast, layers thinner than 5 nm are much harder to peel off, indicating stronger adhesion[19]. For sufficiently thick samples, wet delamination allows BN layers to detach easily from their original sapphire substrates. However, this is not the case for layers transferred onto different substrates. Once a BN layer is transferred, it becomes effectively adhered to the new substrate, preventing further transfers.

To investigate this phenomenon, we conducted experiments where delaminated BN layers were redeposited onto their original sapphire substrates. Exemplary SEM images taken before and after the redeposition process are shown in the Fig 1 a) and b). A significant difference in the sample morphology is observed. The surface of the as-grown sample exhibits wrinkles formed during the growth process[31]. These wrinkles arise due to the mismatch in the thermal expansion coefficients of BN and sapphire[32–34], as the mismatch in thermal expansion coefficients induces compressive stress in the hBN layer, leading to the formation of wrinkles as the material relieves strain.

After the delamination and redeposition procedure, the surface becomes smooth and free of wrinkles. The presence of small crystalline debris[28], which occurs during the growth, in both images before and after delamination (Fig 1) indicates that the BN layer itself remains unchanged, with the only difference being the absence of wrinkles.

It is commonly accepted that delamination of epitaxial boron nitride does not significantly alter its material properties[19,20,31]. To confirm this, we conducted Raman and FTIR measurements, which are non-invasive experimental techniques that do not alter the structural properties of the material.

The analysis of the $E_{2g}$ Raman mode of boron nitride is presented in Fig 1 c). A luminescence background was removed from the spectrum by subtracting a baseline composed of a polynomial. From the fitted Lorentzian peak functions, the as-grown sample exhibited a Raman shift of 1365.0 cm$^{-1}$ and a full width at half maximum (FWHM) of 23.7 cm$^{-1}$. After delamination and redeposition, the Raman shift slightly changed to 1365.5 cm$^{-1}$, and the FWHM remains at a value of 23.7 cm$^{-1}$. The minor change in the Raman shift is most likely attributed to slight variations in the local strain within the crystal lattice. The characteristic values of the $E_{2g}$ mode are nearly unchanged, indicating that the delamination and redeposition processes had almost no impact on the structural properties of the material[20,31].

Fig 1 d) shows the FTIR spectra of the as-grown sample and the same sample after intentional delamination. The prominent peak at ~1367 cm$^{-1}$ corresponds to the $E_{1u}$ phonon mode of sp²-hybridized boron nitride[35], while the high reflectivity below 1000 cm$^{-1}$ is attributed to the sapphire substrate[36]. FTIR data were analyzed by modeling the dynamic dielectric function of the sapphire substrate and BN layer, following the approach outlined in[37]. This method assumes that the materials can be represented as systems of damped harmonic oscillators. From the model, we extracted key parameters, including the oscillator self-energy (phonon energy, corresponding to the peak position), the damping factor (peak width), and the BN layer thickness. An average over seven measurement areas revealed a slight decrease in phonon energy from 1367.6 cm$^{-1}$ for the as-grown sample to 1366.9 cm$^{-1}$ for the delaminated sample, attributed to the reduction of compressive stress during film relaxation and wrinkle removal. Meanwhile, the peak widths (23.8 cm$^{-1}$ and 23.5 cm$^{-1}$) and layer thicknesses (25.3 nm and 25.4 nm) for the as-grown and delaminated layers, respectively, showed minimal variation. These results suggest that the intrinsic properties of the BN layer remain unchanged, with stress relief being the primary effect of wrinkle elimination.

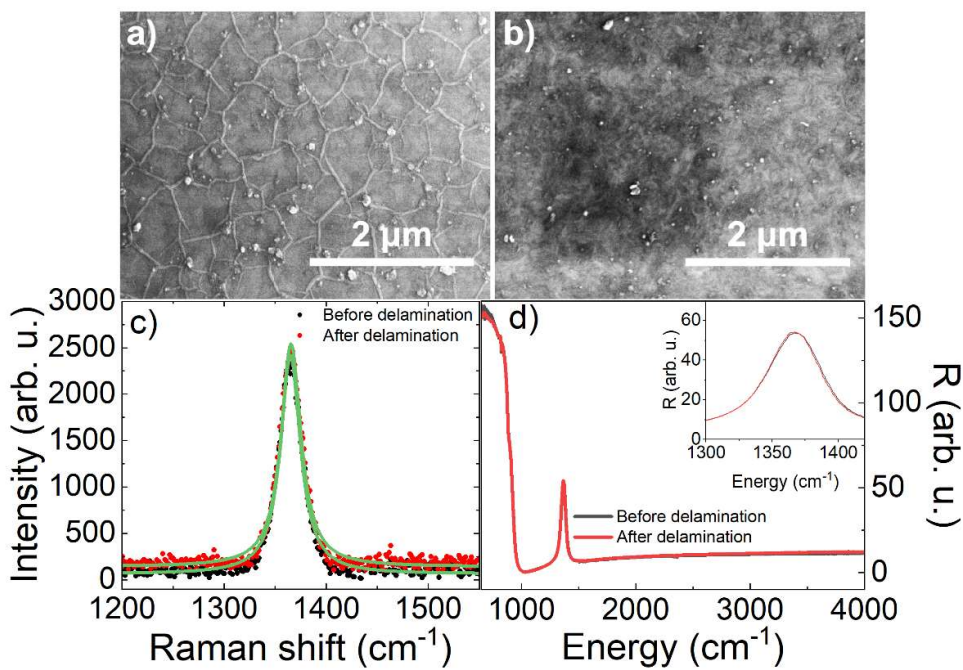

Fig 1. a) SEM image of an as-grown BN layer showing wrinkles formed during cooling after growth due to differences in expansion coefficient between BN and sapphire. b) SEM image of the same BN layer after controlled delamination and redeposition onto the original sapphire substrate, revealing a smooth, wrinkle-free surface. c) Raman spectra of the $E_{2g}$ peaks of BN measured for as-grown and delaminated samples indicate no significant changes among the samples. The green curves represent the fitting of the Lorentzian peak function. d) FTIR spectra comparing the as-grown BN layer with the same sample after delamination, showing no changes in the material's properties. The inset shows a zoom of the hBN $E_{1u}$ vibrational mode peak.

To be able to characterize the electrical and optoelectrical properties of our BN layers, we need to structure the layers using photolithography. Typical results of lithography performed on an as-grown thin BN layer (~30 nm thick) on a sapphire substrate are shown in Fig 2 a). Numerous folds are visible across the entire sample, affecting both regions covered and uncovered by the photoresist. The formation of folds is likely due to liquids penetrating the BN/sapphire interface during the lithographic process. It is important to note that wrinkles are distinct from the folds that develop later as a result of delamination during processing. Wrinkles, which form during cooldown after growth, are sub-micrometer to micrometer in size (see Fig. 1 a)), and not visible under a conventional optical microscope whereas folds, resulting from delamination during processing, are larger (ranging from a few to hundreds of micrometers) and can be observed with an optical microscope.

Fig 2 b) displays a similar BN layer patterned with a Hall bar-shaped photoresist. During the development of the resist, the BN layer delaminates. Folds were present throughout the entire sample, including the BN layer beneath the photoresist. This image was captured without differential interference contrast, resulting in less vivid colors compared to Fig 2 a).

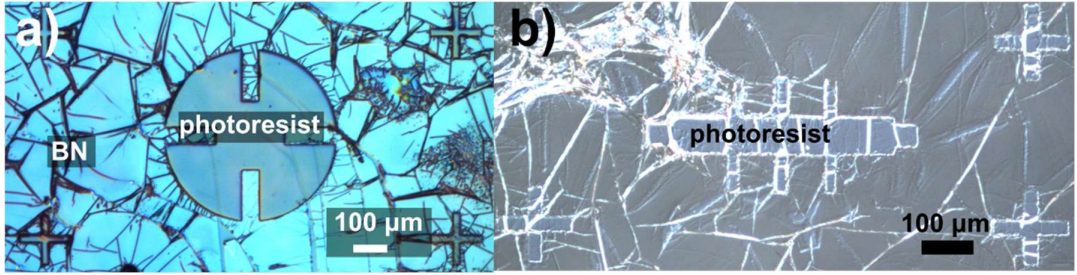

Fig 2. a) Differential interference contrast image of a 30 nm-thick BN layer patterned into a cloverleaf structure with "+"-shaped alignment marks. The image reveals folds across the BN layer, present both on the patterned and surrounding regions, likely caused by liquid penetration at the interface during processing. b) Optical image of a Hall bar-shaped photoresist atop the BN layer. Folds are visible throughout the entire layer, including areas beneath the mask. The image was taken without differential interference contrast, resulting in less vivid colors compared to panel a).

Since the delaminated and redeposited layers could not be further removed by delamination, we propose that removing wrinkles from the BN surface plays a crucial role in preventing the formation of folds during processing. Wrinkles create irregularities on the surface that can facilitate the penetration of water into the interface between the BN layer and the substrate. This penetration increases the likelihood formation of folds, as those shown in Fig 2. By eliminating wrinkles, the surface becomes more flat, reducing the risk of water accessing the BN-sapphire interface and significantly decreasing the chances of uncontrolled delamination during processing and fold formation. Moreover, after delamination, the strain that arises during the cool-down after growth, leading to wrinkle formation, is relaxed[31,37]. Internal forces that assist the delamination process are therefore minimized, making it more difficult to detach the BN layer.

Based on our finding that intentional delamination can produce wrinkle-free epitaxial BN layers, and thus prevent the layer from delamination during lithography processes, we developed a procedure to delaminate and redeposit the layer onto its original substrate, as illustrated in Fig 3 a).

The process begins by gently submerging the sample in a solution of deionized water mixed with a small amount of isopropyl alcohol (IPA) in a water-IPA ratio of 4:1. The addition of IPA reduces the surface tension of the solution, facilitating the penetration of water into the BN/sapphire interface. The sample is positioned at an angle of approximately 20 ° relative to the solution surface to optimize penetration. While this is a typical value, the angle should be experimentally adjusted to the specific characteristics of the shape of sample. As the sample is gradually lowered, the solution enters the interface, allowing the BN layer to detach and float on the water's surface due to surface tension. Near complete detachment, the sample is slowly withdrawn from the solution, enabling the BN layer to reattach to the substrate with near-perfect alignment.

Following this step, residual water trapped at the interface necessitates thorough drying to ensure stability for further processing. The drying procedure begins with heating the sample on a hotplate at 80 °C for 30 minutes. The temperature is then increased to 120 °C, and the sample is heated for an additional 2 hours. Finally, the sample undergoes final drying in a vacuum oven at 200 °C for 12 hours. This procedure ensures the BN layer adheres to the substrate, allowing it to be immersed in water or other polar solvents without the risk of further delamination. Fig 3 b) displays a photograph of a BN sample undergoing intentional delamination. The delaminated section is outlined with a red dashed line. Areas marked with green dashed lines indicate spots where water has penetrated the interface, but the layer has not fully delaminated yet.

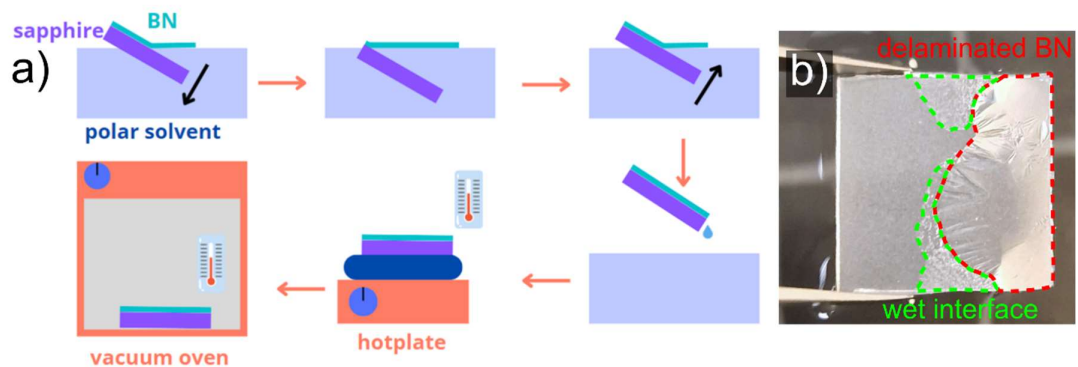

Fig 3. a) Schematic representation of the intentional wet delamination and redeposition process for BN layers. The procedure involves the gradual submersion of a BN sample in a solution of deionized water and IPA, facilitating the detachment of the BN layer from the sapphire substrate. The detached layer is then redeposited onto the substrate. A post-delamination drying step is employed to eliminate residual water from the interface, ensuring structural integrity and preventing further delamination. b) Photograph of an hBN sample undergoing intentional delamination. The delaminated area is outlined with a red dashed curve, while regions where water has penetrated the interface but has not yet detached the entire layer are indicated by green dashed lines.

To demonstrate how this method works for structurization of real devices, we applied the delamination and redeposition procedure to samples with thicknesses of 30 nm and 800 nm. Following these steps, a two-step lithography process was performed. First, RF-ICP etching was employed to define a Hall bar structure in the 30 nm-thick BN layer. Subsequently, metallic contacts were sputtered onto the structure using a gold-palladium (AuPd) alloy, selected for its strong adhesion to BN. An optical image of the fabricated Hall bar device is shown in Fig 4 a).

To remove the metallic layer, the sample was submerged in acetone and subjected to a gentle ultrasonic bath, to aid the penetration of acetone into the photoresist mask. Despite this treatment, the BN Hall bar structures remained intact.

The primary objective of the Hall bar device was to demonstrate our ability to fabricate advanced structures on BN layers while preserving their structural integrity. For this proof of concept, we selected a BN layer that was highly prone to delamination to further emphasize the robustness of our fabrication approach.

As an example of the efficiency of the intentional delamination step we present an exemplary photodetector structure on a 800 nm-thick BN sample. Upon submersion in water, the sample delaminated almost immediately, making further structuring particularly difficult in accordance with what has been reported for layers thicker than 300 nm[17] used for the fabrication of neutron detectors.

The substantial thickness of the BN layer made it an excellent candidate for a photodetector, as it can absorb significant amounts of incoming light. To fabricate the device, we prepared a photolithographic mask in the shape of a photoresistor and sputtered the structure with an AuPd alloy. The metal layer was then removed using a lift-off process described in the Method section. After completing these steps, no signs of degradation or damage to the sample were observed, as shown in Fig 4 b). The photocurrent for the fabricated detector is presented in Fig 4 c). The device was biased at 42 V, and the absorption threshold is approximately at 258 nm. This result highlights the potential of thick BN layers for ultraviolet photodetector applications.

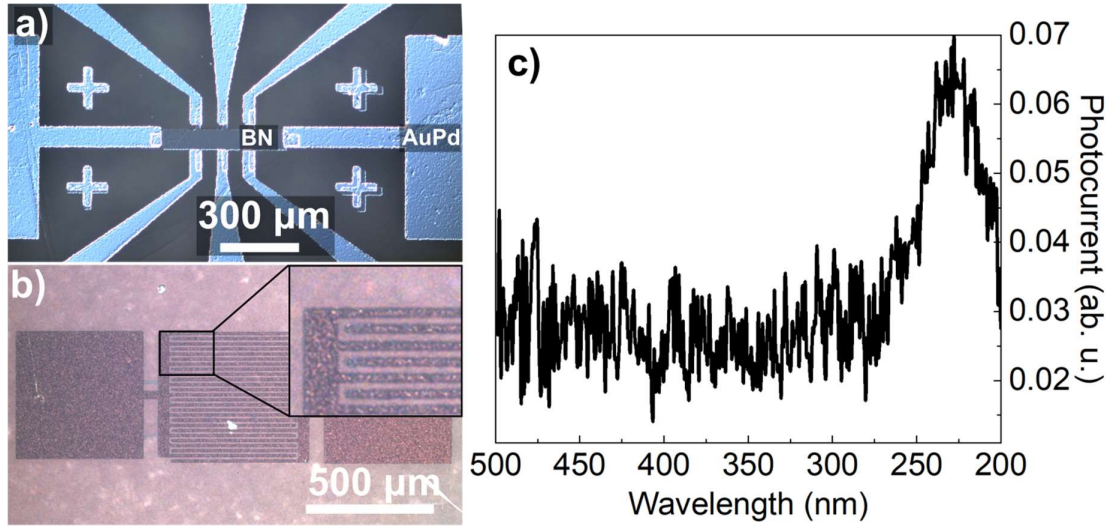

Fig 4. a) Optical image of a Hall bar structure fabricated from a 30 nm-thick epitaxial BN layer with 50 nm-thick AuPd contacts. This demonstrates that our method is suitable for fabricating complex, multilayer devices. b) Optical image of an 800 nm-thick BN layer patterned with 50 nm-thick AuPd contacts. The inset provides a magnified view of the detector structure. c) Photocurrent as a function of wavelength, demonstrating an absorption edge at approximately 258 nm.

The conducted experiments demonstrate that the proposed delamination and redeposition method is highly effective for lithographic structuring of as-grown epitaxial BN layers. The deliberate delamination process successfully eliminates wrinkles from the material, significantly enhancing its suitability for further wet processing on the wafer scale.

## Conclusions

In this work, we presented how intentional delamination of epitaxial BN layers can prevent damage during wet processing. By removing wrinkles that are inherent to high temperature growth and stem from the differences of coefficients of thermal expansion of the sapphire substrate and hBN this method effectively mitigates the risk of further delamination, as the process limits the possibility of water penetration at the BN/sapphire interface, ensuring structural stability. We successfully applied the intentional delamination technique to fabricate a Hall bar structure. The process involved an etching step followed by the deposition of metallic contacts. Furthermore, the method proved effective for thicker BN layers (800 nm), which were used to fabricate a photodetector. This approach provides a straightforward and reliable solution for improving the quality and processability of epitaxial BN layers. It offers a valuable tool for researchers exploring BN-based technologies and has the potential to enable industrial applications in advanced optoelectronic and semiconductor devices. The processing of thicker layers is especially important for the realization of neutron detectors for which the thickness of the layers is one of the crucial parameters.

## Acknowledgements

This work was supported by the Polish National Science Centre, Poland, under decisions 2020/39/D/ST7/02811, 2021/41/N/ST7/04326, 2021/41/N/ST3/03579, 2022/47/B/ST5/03314.